\documentclass[conference]{IEEEtran}
\usepackage{cite}
\usepackage{amsmath,amssymb,amsfonts}
\usepackage{algorithmic}
\usepackage{graphicx}
\usepackage{textcomp}
\usepackage{xcolor}
\usepackage{booktabs}
\usepackage{multirow}
\usepackage[hidelinks]{hyperref} % 禁用彩色链接
\usepackage{algorithm}
\usepackage{algorithmic}
\usepackage{graphicx}
\usepackage{float}
\usepackage{arydshln}
\usepackage{wrapfig}
\usepackage{lettrine}
\usepackage{subcaption}
\usepackage{caption}
\usepackage{etoolbox}
\usepackage{pifont}
\usepackage{marvosym}  % 或者
\usepackage{ifsym}     % 也可以用于定义\Letter符号
\usepackage[utf8]{inputenc}
\usepackage[utf8]{inputenc}
\usepackage[T1]{fontenc}

\DeclareUnicodeCharacter{FF0C}{,} % 替换中文逗号为英文逗号
\DeclareUnicodeCharacter{2013}{--} % 替换长破折号为双破折号
\begin{document}
\title{Adaptive Hybrid FFT: A Novel Pipeline and Memory-Based Architecture for Radix-\(2^k\) FFT in Large Size Processing}
\author{
    \IEEEauthorblockN{
        Fangyu Zhao\IEEEauthorrefmark{1}\IEEEauthorrefmark{2}, 
        Chunhua Xiao\textsuperscript{\Letter}\IEEEauthorrefmark{1}\IEEEauthorrefmark{2}, 
        Zhiguo Wang\IEEEauthorrefmark{3}, 
        Xiaohua Du\IEEEauthorrefmark{3}, 
        Bo Dong \IEEEauthorrefmark{3}
    }
    \IEEEauthorblockA{
        \IEEEauthorrefmark{1}College of Computer Science, Chongqing University, Chongqing, China\\
        \IEEEauthorrefmark{2}Key Laboratory of Dependable Service Computing in Cyber Physical Society, Ministry of Education, China
    }
    \IEEEauthorblockA{
        \IEEEauthorrefmark{3}Sichuan Huacun Zhigu Technology Co., Ltd., Chengdu, China
    }
    \IEEEauthorblockA{
{\textit{xiaochunhua}@cqu.edu.cn},{\textit{zhaofangyu}@stu.cqu.edu.cn},\{\textit{duxiaohua, dongbo, wangzhiguo\}@tgstor.cn，
}
}
    \IEEEauthorblockA{
       \textsuperscript{\Letter} Corresponding Author
    }
}
\maketitle
\textbf{Disclaimer:}This work has been submitted to the IEEE for possible publication.
Copyright may be transferred without notice, after which this version may no longer be accessible.
 
\begin{abstract}
In the field of digital signal processing, the fast Fourier transform (FFT) is a fundamental algorithm, with its processors being implemented using either the pipelined architecture, well-known for high-throughput applications but weak in hardware utilization, or the memory-based architecture, designed for area-constrained scenarios but failing to meet stringent throughput requirements. Therefore, we propose an adaptive hybrid FFT, which leverages the strengths of both pipelined and memory-based architectures. In this paper, we propose an adaptive hybrid FFT processor that combines the advantages of both architectures, and it has the following features. First, a set of radix-\(2^k\)multi-path delay commutators (MDC) units are developed to support high-performance large-size processing. Second, a conflict-free memory access scheme is formulated to ensure a continuous data flow without data contention. Third, We demonstrate the existence of a series of bit-dimension permutations for reordering input data, satisfying the generalized constraints of variable-length, high-radix, and any level of parallelism for wide adaptivity. Furthermore, the proposed FFT processor has been implemented on a field-programmable gate array (FPGA). As a result, the proposed work outperforms conventional memory-based FFT processors by requiring fewer computation cycles. It achieves higher hardware utilization than pipelined FFT architectures, making it suitable for highly demanding applications.\\
\end{abstract}


\section{Introduction}
The fast Fourier transform (FFT) algorithm is widely used in the field of digital signal processing, including wireless communication, image processing, and radar signal processing. Nowadays, the demand for handling large-size data has been ever-increasing.\\
\indent Over the past decades, studies on FFT hardware implementation have primarily focused on two solutions: the pipeline architecture \cite{Garrido2013,groginsky1970pipeline,ingemarsson2018sff} and the memory-based architecture \cite{tsai2011generalized,kaya2023memory,wang2020scheduling,xia2017memory}. The pipelined hardware architecture is employed in high-throughput real-time applications due to its fully horizontal continuous flow. Conversely, the memory-based architecture is well-suited for area-constrained scenarios, as it allows a set of arithmetic hardware to be adapted to variable sizes, achieving better hardware utilization. \\
\indent Despite the aforementioned advantages, the pipeline architecture can result in low hardware utilization, as a significant portion of the hardware remains idle when performing small-size operations on an FFT processor designed for large FFT-size applications.

In contrast, memory-based architecture serves as an alternative approach for efficient resource design with only one set of arithmetic processing units. However, the computational cycle is increased since larger FFT sizes must be iterated multiple times over the same arithmetic units, which is detrimental to real-time processing. Therefore, to circumvent this problem, some studies\cite{tsai2011generalized,wang2020scheduling,xia2017memory} have focused on utilizing high-radix multipath delay commutators (MDC) as core processing units, enhancing computational efficiency. Moreover, the radix of the algorithm greatly influences the complexity of computation, high radix will reduce the overall computational complexity by decreasing the number of stages. 

Additionally, researchers have focused on improving the memory-based FFT architecture. For instance, considering memory dissipation, utilizing only N-size memory for N-point FFT calculations is highly feasible.  Conflict-free memory addressing schemes become crucial for achieving this design with a theoretical minimum memory requirement while eliminating concurrent read and write collisions. Specifically, the study in \cite{ma2015novel} proposes a conflict-free memory addressing scheme for radix-2 real FFT. In \cite{kaya2023memory}, a perfect bit permutation of the memory circular counter is introduced to implement an in-place strategy for the radix-2 algorithm. However, the aforementioned approaches are strictly confined to radix-2 FFT processors, which do not take into account radix-$2^k$ butterfly units.

Apart from conflict-free address schemes, permutation circuits \cite{kaya2023memory}\cite{tsai2011generalized}\cite{824693}\cite{Garrido2017}, which are employed to reorder data between the memory and a set of arithmetic units, are also essential components. Data layout transformation is applied to ensure the correct sequence order of FFT calculations. In \cite{kaya2023memory}, the permutation circuits comprising four multiplexers and four delay buffers for radix-2 calculation are discussed. However, the limitation also lies in the lack of comprehensive support for radix-$2^k$ calculation.
 
In this paper, we propose an adaptive hybrid FFT processor capable of reconfiguring both pipeline and memory-based architectures. This architecture supports FFT sizes up to 512K. To accelerate computational speed, we utilize multiple processing elements (PEs) consisting of radix-$2^k$ multi-path delay commutators. Additionally, the proposed architecture implements efficient conflict-free memory access schemes for each reconfigurable architecture, ensuring a continuous data flow without underlying contention issues. Furthermore, we extend the permutation circuit described in existing work \cite{kaya2023memory} and demonstrate the existence of a series of bit-dimension permutations for reordering input data, meeting the broader requirements under any power-of-two FFT size, any level of parallelism, and any radix.

 Our main contributions can be summarized as follows:
\begin{itemize}
    \item We propose an adaptive hybrid FFT processor architecture that supports both pipeline and memory-based approaches for large 512K sizes. 
    \item We introduce conflict-free memory access techniques that enable continuous-flow FFT computations through fully in-place memory transformations at each iteration. This approach is customized for different reconfigurable architecture targets.
    \item We develop a comprehensive bit permutation methodology that accommodates diverse hardware constraints, such as varying FFT lengths, radices, and levels of parallelism.
\end{itemize} 

\indent The paper is organized as follows: Section \ref{sec:sectionII} covers background information. Section \ref{sec:sectionIII} describes the top-level design. Section \ref{sec:sectionIV} details the adaptive hybrid FFT processors. Section \ref{sec:sectionV} compares the proposed FFT with prior state-of-the-art solutions. Finally, Section \ref{sec:sectionVI} concludes the paper.

\section{Background}\label{sec:sectionII}
In this section, we provide an overview of the fundamental techniques relevant to our proposed architecture.
\subsection{Bit-Dimension Permutation}
 Bit-Dimension permutation\cite{garrido2019optimum} is highly amenable for reordering samples. This technique rearranges data by shuffling the primitive order of $n$ bits. The detailed process is outlined below: the sample index of an $N$-point ($N = 2^n$) sequence is calculated as $\mathcal{P}_{\text{index}} = \sum_{i=0}^{n-1} x_{i} 2^{i}$, with $x_{i}$ representing the $i$-th bit dimension and $x_{i} \in \{0, 1\}$. It can also be simplified as:
 \begin{equation}
 \mathcal{P}_{\text{index}} \equiv x_{n-1} x_{n-2} \ldots x_{0}
 \end{equation}
 where symbol $\equiv$ denotes the binary representation of the index. Meanwhile, bit-dimension permutation, $\sigma$, is defined to coordinate every bit-dimension position by applying the same transferring rule\cite{Fraser1976}, which can be expressed as the following equation:
 \begin{equation}
\sigma (x_{n-1} x_{n-2} \ldots x_0) = x_{\sigma(n-1)} x_{\sigma(n-2)} \ldots x_{\sigma(0)}
 \end{equation}
where $x_{\sigma(n-1)}$ represents the new bit on $(n-1)$-th dimension after the permutation.
For instance, $\sigma(x_{3}, x_{2}, x_{1}, x_{0}) = (x_{3}, x_{0}, x_{1}, x_{2})$ refers to reversing the last three bits and concatenating them with the most significant bit\cite{kaya2023memory}. 
\subsection{\texorpdfstring{Radix-$2^5$ Algorithm}{Radix-2k Algorithm}}
To circumvent the problem that the hardware implementation of radix-2 involves many complex multipliers and radix-4/8 require more hardware complexity for implementing the butterfly unit, the radix-$2^k$ algorithm is proposed. This algorithm simultaneously achieves a simple butterfly structure and minimizes the number of complex multipliers \cite{cortes2009radix}. And the radix-$2^5$ algorithm can be derived from discrete Fourier Transform(DFT) of N-length by using the following equation:
\begin{equation}
X(k) = \sum_{n=0}^{N-1} x(n) W_N^{nk}\label{eq:DFT}
\end{equation}
where the $n$ is the time index, $k$ is the frequency index and $W_N$ is the twiddle factor.To induce the expression of radix-$2^5$ algorithm, a 6-dimension linear map is applied as follows:
 \begin{equation}
\begin{cases}
n = \frac{N}{2} n_1 + \frac{N}{4} n_2 + \frac{N}{8} n_3 + \frac{N}{16} n_4 + \frac{N}{32} n_5 + n_6 \\
n_1, n_2, n_3, n_4, n_5 = 0, 1 \\
n_6 = 0, 1, \ldots, \frac{N}{32} - 1 \\
k = k_1 + 2k_2 + 4k_3 + 8k_4 + 16k_5 + 32k_6\\
k_1, k_2, k_3, k_4, k_5 = 0, 1 \\
k_6 = 0, 1, \ldots, \frac{N}{32} - 1
\end{cases}
\end{equation}
Substituting this mapping method into the Equation (\ref{eq:DFT}) results in the following:
\begin{small}
\begin{equation}
\begin{aligned}
X&(k_1 + 2k_2 + 4k_3 + 8k_4 + 16k_5 + 32k_6)\\
&= \sum_{n_6=0}^{\frac{N}{32}-1} \sum_{n_5=0}^{1} \sum_{n_4=0}^{1} \sum_{n_3=0}^{1} \sum_{n_2=0}^{1} \sum_{n_1=0}^{1} \\
&\times x\left( \frac{N}{2} n_1 + \frac{N}{4} n_2 + \frac{N}{8} n_3 + \frac{N}{16} n_4 + \frac{N}{32} n_5 + n_6 \right) W_N^{nk}
\end{aligned}
\end{equation}
\end{small}
And the twiddle factor decomposition is expressed as:
\begin{equation}
W_N^{nk} = W_N^{
\begin{aligned}
&\left(\scriptstyle
\frac{N}{2} n_1 + \frac{N}{4} n_2 + \frac{N}{8} n_3 + \frac{N}{16} n_4 + \frac{N}{32} n_5 + n_6 
\right) \\
&\left(\scriptstyle
k_1 + 2k_2 + 4k_3 + 8k_4 + 16k_5 + 32k_6
\right)
\end{aligned}
}
\label{eq:tfd}
\end{equation}
Observing Equation (\ref{eq:tfd}), the decomposition results are varied with regard to the common factor algorithm \cite{cho2013high}. Thus, the twiddle factor in each stage of radix-$2^5$ can be expressed in various forms. 

\section{Top-Level Hybrid-FFT Processor Description}\label{sec:sectionIII}
\begin{figure}[ht!]
    \centering
    \includegraphics[width=0.4\textwidth]{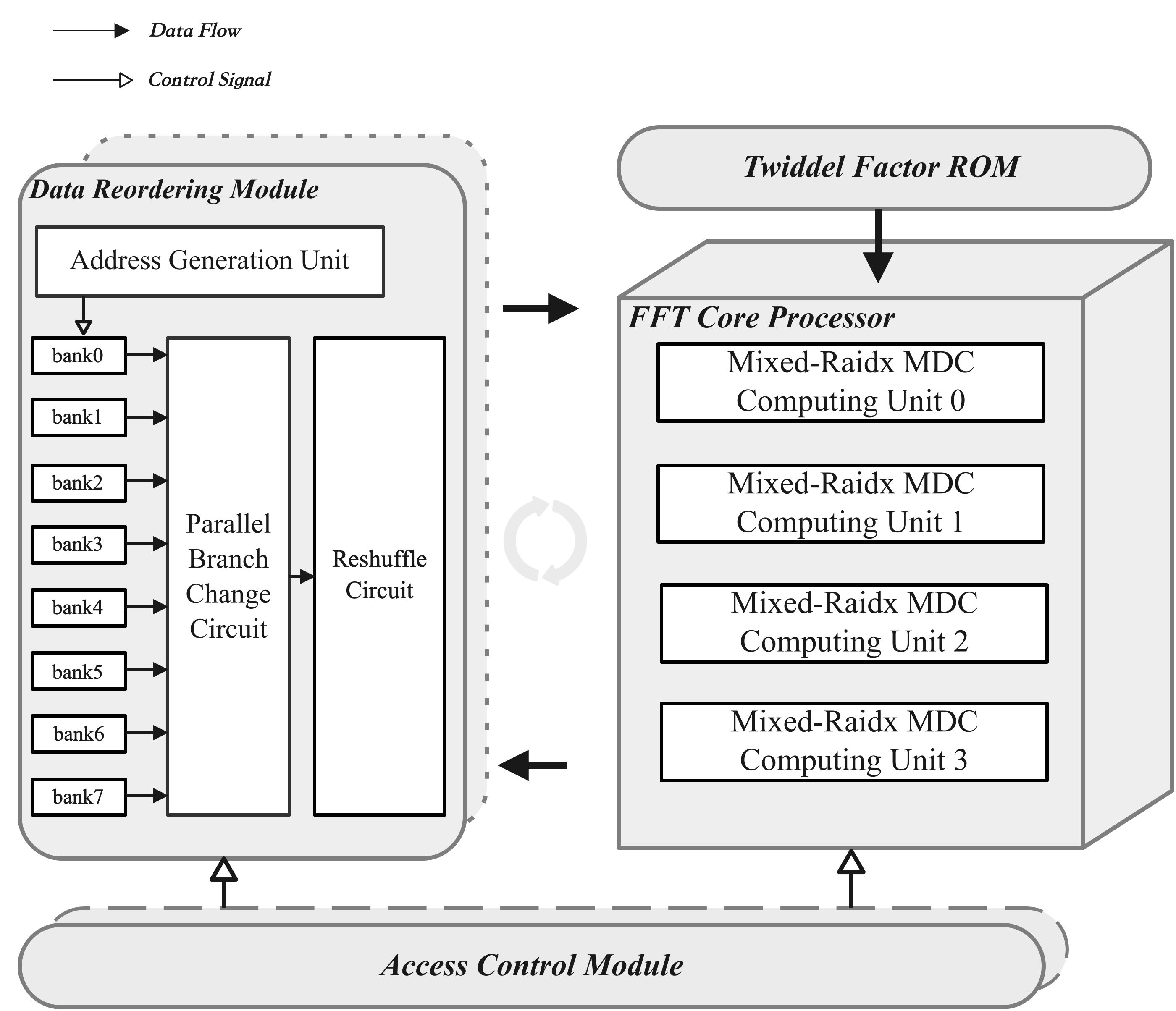}
    \caption{Top-level of the proposed FFT architecture for 512K sizes}
    \label{fig:overview}
\end{figure}

\begin{figure}[ht] % 确保图片在当前位置浮动
    \centering
    \includegraphics[width=0.4\textwidth]{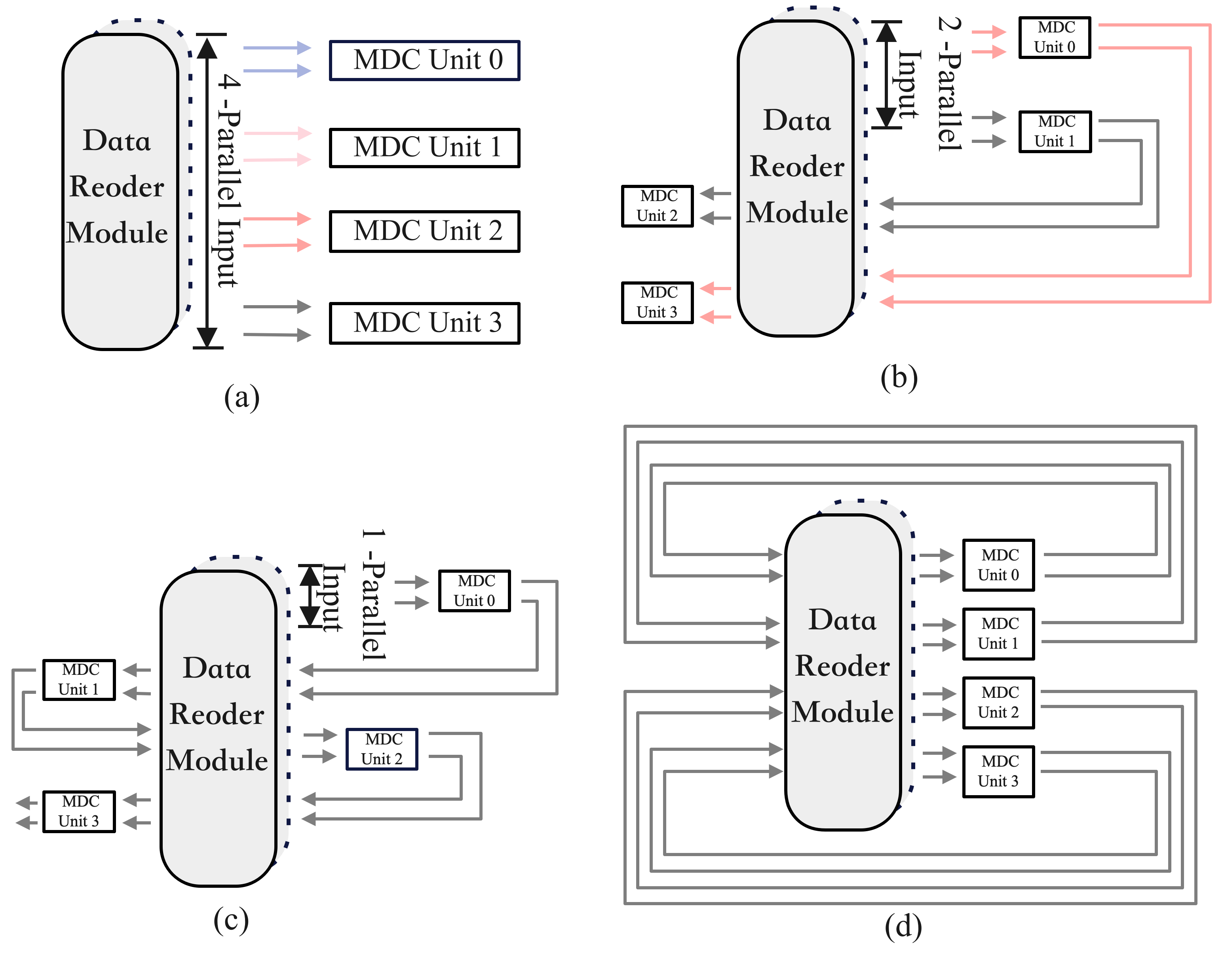} % 调整图像宽度
    \caption{Computational mode of the FFT processor}
    \label{fig:work_mode}
\end{figure}

The top level of the proposed FFT architecture is shown in Fig.\ref{fig:overview}. It consists of two essential modules: the data reordering module and the FFT core processor. The FFT core processor is composed of a set of identical radix-$2^k$ MDC units as detailed in Fig.\ref{fig:sample-image}. The data reordering module includes memory banks, a parallel branch permutation circuit, and a reshuffle circuit.
The proposed FFT architecture employs $P$ parallelism, with samples stored in $2P$ memory banks. Our architecture is designed to support 512K sizes with radix-$2^5$, which requires four MDC units and eight banks. These memory banks not only store the FFT data but also permute indices via the address generation unit, denoted as $\sigma_1$. The parallel branch permutation circuit represents another permutation, $\sigma_2$, while the reshuffle circuit performs the final permutation, $\sigma_3$. The calculation of the permutations $\sigma_1$, $\sigma_2$, and $\sigma_3$ results in the combined permutation $\sigma = \sigma_3 \circ \sigma_2 \circ \sigma_1$, which ensures the correct input order to the MDC unit.
Details will be further discussed in Section \ref{sec:sectionIII-B}.\\
\indent As illustrated in Fig.\ref{fig:work_mode}, the proposed FFT architecture can be categorized into two types: pipeline mode and memory-based mode. Figures \ref{fig:work_mode}(a)-(c) depict three scenarios under the pipeline mode with different levels of parallelism, capable of handling different $P$ batches of $N$ points concurrently, highlighted in different colors. Each memory has $N/2$ addresses. Fig.\ref{fig:work_mode}(a) is designed for calculating points where $N \in [2, 2^{k}]$, requiring only one computation stage. Fig.\ref{fig:work_mode}(b) is suitable for calculating points where $N \in [2^{k+1}, 2^{2k}]$. Lastly, Fig.\ref{fig:work_mode}(c) is applicable for calculating points where $N \in [2^{2k+1}, 2^{4k}]$. Fig.\ref{fig:work_mode}(d) depicts the data-flow for a memory-based working process, where $N$ is limited to $[2^{k}, 2^{3k}]$. One batch of $N$ points is executed with $P=4$ parallelism. Each memory has $N/2P$ addresses. The distinctive difference between these two modes is the storage location of intermediate results. In pipeline mode, the results are placed into neighboring memories, which guarantees continuous flow access. On the contrary, in memory-based mode, the results are stored in the same location of the memory where they were previously read in the last stage.
\section{Elaborated Hardware Design of Hybrid-FFT Processor}\label{sec:sectionIV}
This section introduces the single MDC structure for implementing the radix-$2^5$ algorithm and our solution for accurate FFT calculations under the proposed architecture, with an illustrative example. Additionally, conflict-free memory access methods are discussed to ensure continuous operation.
\subsection{\texorpdfstring{MDC structure with modified radix-$2^5$ Algorithm}{Radix-2k Algorithm}}
\begin{figure}[ht!]
    \centering
    \includegraphics[width=0.4\textwidth, height=0.4\textheight]{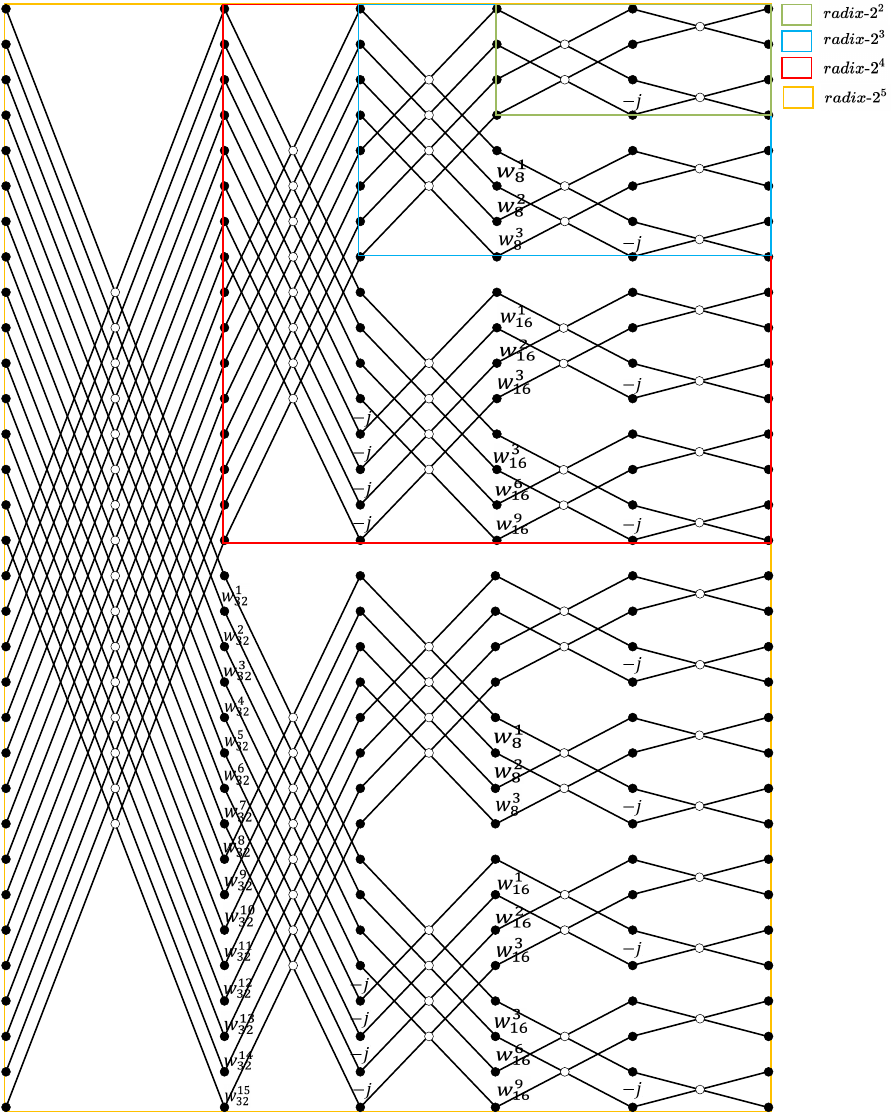}
    \caption{Signal Flow graph of the radix-$2^5/2^4/2^3/2^2$ Algorithm}
    \label{fig:high_radix_32_point}
\end{figure}

\begin{figure}[ht]
    \centering
    \includegraphics[width=0.5\textwidth]{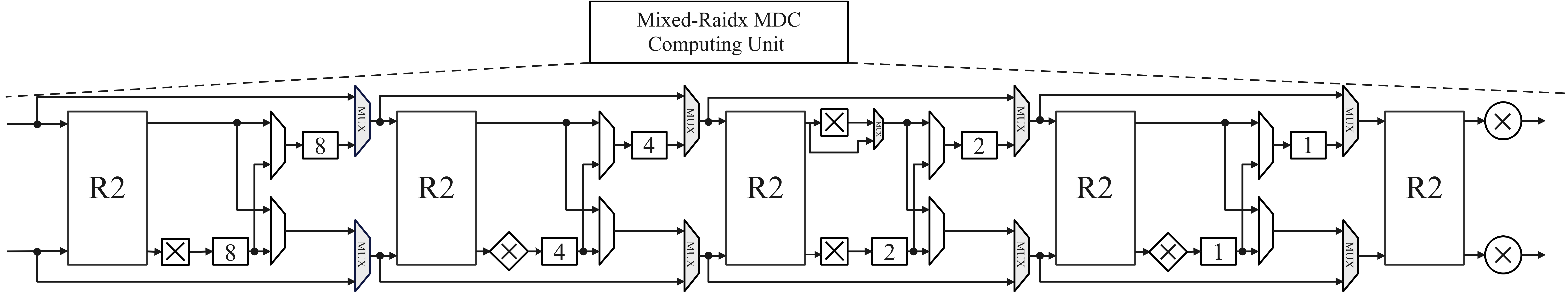} % 替换为你的图片路径
    \caption{Proposed MDC architecture for radix-$2^5/2^4/2^3/2^2$ computation} % 图片说明
    \label{fig:sample-image} % 标签，便于引用
\end{figure}

The proposed MDC design includes architectures to calculate radix-$2^5$, radix-$2^4$, radix-$2^3$, and radix-$2^2$ algorithm. To achieve a hardware-efficient design, it is intended that the same set of hardware should be utilized for mixed-radix computation scenarios. However, this goal is unachievable because the structure of twiddle factor multiplication varies with radix $k$ \cite{yang2016high}. Nevertheless, a study \cite{Garrido2013} adopts a modified radix-$2^3$/$2^4$ algorithm to enable a scalable and reusable MDC structure. Likewise, inspired by this work, we extend the approach to account for radix-$2^5$. Additionally, we classified rotation into three categories: constant (C), non-trivial (NT), and trivial (T), as shown in the hardware implementation in Fig.\ref{fig:sample-image}. According to (\ref{eq:tfd}), the modified radix-$2^5$ algorithm is expressed as:
\begin{equation}
\begin{aligned}
&W_N^{nk} = W_N^{
\begin{aligned}
&\left(\scriptstyle
\frac{N}{2} n_1 + \frac{N}{4} n_2 + \frac{N}{8} n_3 + \frac{N}{16} n_4 + \frac{N}{32} n_5 + n_6 
\right) \\
&\left(\scriptstyle
k_1 + 2k_2 + 4k_3 + 8k_4 + 16k_5 + 32k_6
\right)
\end{aligned}
} \\
&= \underbrace{(-1)^{n_1k_1}}_{\text{Stage 1 BU}} \overbrace{W_{32}^{k_1(8n_2 + 4n_3 + 2n_4 + n_5)}}^{\text{Stage 1 TF}} \underbrace{(-1)^{n_2k_2}}_{\text{Stage 2 BU}} \overbrace{(-j)^{n_3k_2}}^{\text{Stage 2 TF}} \\
&\times \underbrace{(-1)^{n_3k_3}}_{\text{Stage 3 BU}} \overbrace{W_{32}^{2(k_2 + 2k_3)(2n_4 + n_5)}}^{\text{Stage 3 TF}} \underbrace{(-1)^{n_4k_4}}_{\text{Stage 4 BU}} \overbrace{(-j)^{n_5k_4}}^{\text{Stage 4 TF}} \\
&\times \underbrace{(-1)^{n_5k_5}}_{\text{Stage 5 BU}} \overbrace{W_N^{(k_1 + 2k_2 + 4k_3 + 8k_4 + 16k_5)}}^{\text{Stage 5 TF}} W_{\frac{N}{32}}^{n_6k_6}
\end{aligned} 
\end{equation}
where $W_{32}$ is the constant twiddle factor preliminarily stored in ROM, $W_{N}$ is the non-trivial twiddle factor with respect to $N$, and $-j$ is considered a trivial factor. Table \ref{your_table_label_IVf} lists the twiddle factor properties of the radix-$2^2$, radix-$2^3$, radix-$2^4$, and radix-$2^5$. Reviewing Table \ref{your_table_label_IVf} row by row, we note that the properties of the twiddle factor remain consistent across different radix-$2^k$. For instance, the second stage of radix-$2^5$ requires trivial rotators, which is in accordance with the first stage of radix-$2^4$. This consistency allows the reuse of the twiddle factor multiplier across various radices, thereby achieving a hardware-efficient design. Additionally, Fig.\ref{fig:high_radix_32_point} shows the signal flow graph of the radix-$2^5/2^4/2^3/2^2$ algorithm. It can be seen that the coefficient associated with radix-$2^5$ is also compatible with other lower radices. Due to the symmetric distribution of trivial twiddle factors $(-j)$ shown in Fig.\ref{fig:high_radix_32_point}, both the hardware structure and the control mechanism for trivial multipliers can be applied concurrently to various radices. Therefore, Fig.\ref{fig:sample-image} depicts the 32-point radix-$2^5$ MDC architecture, which consists of radix-2 butterflies, constant multipliers, non-trivial multipliers, trivial multipliers, and shuffling structures. The lengths of the buffers are indicated by the number. If the radix-$2^4/2^3/2^2/2$ operation is required, the input signals will bypass the preceding butterflies via multiplexers to enable mixed-radix calculations in a single MDC. Moreover, the MDC structure shown in  Fig.\ref{fig:sample-image} corresponds to the computing unit of FFT Core Processors of Fig.\ref{fig:overview}.
\begin{table}[ht!]
\caption{Properties of the Twiddle Factor}
\label{your_table_label_IVf}
\resizebox{0.45\textwidth}{!}{
\begin{tabular}{cc|cc|cc|cc}
\hline
\multicolumn{2}{c|}{radix-$2^5$}            & \multicolumn{2}{c|}{radix-$2^4$}            & \multicolumn{2}{c|}{radix-$2^3$}            & \multicolumn{2}{c}{radix-$2^2$}            \\ \hline\hline
\multicolumn{1}{l:}{Stage} & \multicolumn{1}{l|}{Property} & \multicolumn{1}{l:}{Stage} & \multicolumn{1}{l|}{Property} & \multicolumn{1}{l:}{Stage} & \multicolumn{1}{l|}{Property} & \multicolumn{1}{c:}{Stage} & \multicolumn{1}{l}{Property} \\ \hline
\multicolumn{1}{c:}{1}     & $W_{32}$                            & \multicolumn{1}{c:}{-}     & -                           & \multicolumn{1}{c:}{-}     & -                         & \multicolumn{1}{c:}{-}     & -                             \\  \hline
\multicolumn{1}{c:}{2}     & T                             & \multicolumn{1}{c:}{1}     & T                       & \multicolumn{1}{c:}{-}     & -                             & \multicolumn{1}{c:}{-}     & -                           \\  \hline
\multicolumn{1}{c:}{3}     & $W_{32}$                             & \multicolumn{1}{c:}{2}     & $W_{16}$                             & \multicolumn{1}{c:}{1}     & $W_{8}$                             & \multicolumn{1}{c:}{-}     & -                              \\  \hline
\multicolumn{1}{c:}{4}     & T                             & \multicolumn{1}{c:}{3}     & T                             & \multicolumn{1}{c:}{2}     & T                             & \multicolumn{1}{c:}{1}     & T                             \\  \hline
\multicolumn{1}{c:}{5}     & NT                            & \multicolumn{1}{c:}{4}     & NT                            & \multicolumn{1}{c:}{3}     & NT                            & \multicolumn{1}{c:}{2}     & NT                            \\ \hline
\end{tabular}
}
\end{table}

\subsection{Data Reordering Strategy via Bit Dimension Permutation}\label{sec:sectionIII-B}
Without loss of generality, the radix-$2^{k}$ FFT algorithm computes an $N$-point FFT ($N = 2^{n}$) in $S = \lceil \frac{n}{k} \rceil$ stages, where $\lceil \cdot \rceil$ denotes the ceiling operation. The radix number $k_{s}$ for each stage $s \in \{1, 2\ldots, S\}$ is given by:
\begin{equation}
k_1 = n - k(S - 1), \quad k_2 = \cdots = k_{S}= k.
\end{equation}
At each stage $s$, a radix-$2^{k_{s}}$ MDC unit computes a total of $N/2^{k_{s}}$ butterfly blocks under $P=1$ parallelism, each containing $2^{k_{s}}$ samples. Besides, the sample indices associated in each block must be arranged in a specific order, forming the MDC input for stage $s$. Every block is represented as:
\begin{equation}
B_{s}^{I}(m) = \begin{pmatrix}
b_{0,2^{k_{s}-1}-1}  &\ldots & b_{0,1} & b_{0,0} \\
b_{1,2^{k_{s}-1}-1} &\ldots & b_{1,1} & b_{1,0} \\
\end{pmatrix}_{2\times 2^{k_{s}-1}} \label{eq:block}
\end{equation}
where samples from the first and second rows of $B_{s}^{I}(m)$ are transformed to the MDC lower and upper inputs, respectively. The element $b_{r,c}$ represents the row and column indices of the matrix $B_{s}^{I}(m)$, with $r \in \{0,1\}$ and $c \in \{0,1,\ldots,2^{k_{s}-1}\}$. Define $b_{0,0} = x(m)$, where $m$ is an index of the time-domain sample sequence $x$. The remaining elements in $B_{s}^{I}(m)$ must follow the relation $b_{r,c} = x(m+d)$ in terms of operating on correct pairs of sample within MDC. The variable $d$ is given as:
\begin{equation}
d = \frac{N}{2^{\sum_{u=0}^{s}k_{u}}}(2^{k_{s}-1}\times r+c) \label{eq:d}
\end{equation}
From the aforementioned analysis, it should be noted that a total of $N/2^{k_{s}}$ butterfly blocks arrives in consecutive clock cycles and is sequentially fed into the MDC, those of which are differentiated by the upper-right leading element $b_{0,0}=x(m)$. Once the sample index $m$ is determined, the other indices can be derived from~(\ref{eq:d}). Furthermore, considering an additional condition $k_{0}=0$, the range of $m$ in different stages $s$ is concluded as follows:
\begin{equation}
m\in\begin{cases}
\bigcup\limits_{v=0}^{2^{\sum_{u=0}^{s-1} k_u} - 1} \{v \varepsilon_{s-1} : 1 : v \varepsilon_{s-1} + \varepsilon_{s}\} & s\neq S\\
\{0:2^{k_s}:N\} & s=S
\end{cases} \label{eq:m}
\end{equation}
and with incremental symbol $\varepsilon_{s}$ defined as:
\begin{equation}
\varepsilon_{s} = {N}/{2^{\sum_{u=0}^{s}k_{u}}} \label{eq:eplison}
\end{equation}
where brace notation \{a:i:b\} delineates the sequence increase from integer $a$ to $b$ by step $i$. An elaborate example is provided to specify the above theoretical formulas. We use a radix-$2^5$ MDC to process a 4096-point FFT, with mixed-radix for three stages: $k_{1}=2$, $k_{2}=5$, and $k_{3}=5$, respectively. At the first ($s$=1) stage $d=1024(2r+c)$ is obtained from~(\ref{eq:d}). The first MDC received a time-domain sample block led by index $m=0$, arranged as follows:
\begin{equation}
\scalebox{0.8}{$B_{1}^{I}(0) = \begin{pmatrix}
x(1024) & x(0) \\
x(3072) & x(2048)  \\
\end{pmatrix}_{2\times 2}$} \label{eq:block_i_1}
\end{equation}
with $m \in [0:1:1024)$ from~(\ref{eq:m}). Likewise, at the second stage, $d=32(2r+c)$, a sample block led by index $m=0$ will be input into the second MDC, and:
\begin{equation}
\scalebox{0.8}{$B_{2}^{I}(0)  = \begin{pmatrix}
x(480) & x(448)\ldots &x(32)& x(0) \\
x(992) & x(960)\ldots &x(544)& x(512)  \\
\end{pmatrix}_{2\times 16}$} \label{eq:block_2}
\end{equation}
with merged interval $m \in [0:1:32) \cup [1024:1:1056)\cup [2048:1:2080) \cup [3072:1:3104)$. Finally, at the last stage, the parameter $d$ is reduced to 1. The matrix $B_{3}^{I}(0)$ can be derived by applying formulas (\ref{eq:d}) to (\ref{eq:eplison}) again. When sample blocks $B_{s}^{I}(m)$ are delivered to MDC, the internal reordering circuit naturally adjusts the element indices within each block, and the output results are given as:
\begin{equation}
B_{s}^{O}(m) = \begin{pmatrix}
b_{1,2^{k_{s}-1}-2}\dots &b_{1,0}  & b_{0,2^{k_{s}-1}-2} \ldots b_{0,0} \\
b_{1,2^{k_{s}-1}-1}\dots &b_{1,1}  &b_{0,2^{k_{s}-1}-1} \ldots b_{0,1} \\
\end{pmatrix} \label{eq:block_o}
\end{equation}
For example, $B_{1}^{O}(0)=$ \scalebox{0.8}{$\begin{pmatrix} x(2048) & x(0) \\ x(3072) & x(1024) \end{pmatrix}$}. Intermediate results will be streamed to the second MDC row when the computation of the first MDC row is completed. However, $B_{1}^{O}(0)$ is apparently inconsistent with $B_{2}^{I}(0)$ as mentioned in~(\ref{eq:block_2}). Similarly, incompatibility still occurs in other stages. Hence, the problem arises when intermediate results are transferred to a neighboring MDC in the proposed FFT processor.

With the aim of solving the problem, we propose an input data reordering strategy utilizing bit dimension permutation to circumvent the issue of neighboring MDC straightforwardly receiving the output.
Specifically, an input data reordering strategy is achieved by read/write address generation, parallel branch permutation, and composition of reshuffle circuit as shown in Fig.\ref{fig:overview}, both of which facilitate reordering scrambled output results into the correct input form for the neighboring MDC. We are greatly inspired by previous work \cite{kaya2023memory} but with further extension to consider radix-$2^{k}$ FFT algorithm.

The samples are stored across $2P$ memory banks. Initially, samples are stored in natural order; that is, the $q$-th memory bank stores data with sequential indices $q(N/2P), q(N/2P) + 1, \ldots, (q + 1)(N/2P) - 1$. Assume the sample with index $a$ is stored at address $a$. Therefore, the position of each index can be expressed as:
\begin{equation}
\mathcal{P}_0 \equiv \underbrace{a_{n-1}, a_{n-2}, \ldots, a_{n-p}}_{\text{parallel (memory)}} \mid \underbrace{a_{n-p-1}, \ldots, a_{0}}_{\text{serial (address)}} \label{eq:block_p}
\end{equation}
where \( p = \log_{2} 2P \), the parallel sequence \(\left( a_{n-1}, \cdots, a_{n-p} \right)\) indicates the memory identifier and the serial sequence determines the specific address in \(\sum_{i=0}^{p-1} 2^{i} a_{n-p+i}\)-th memory.

The bit permutation from the $s$-th stage output $B_{s}^{O}$ to the next stage input $B_{s+1}^{I}$ in an arbitrary radix-$k$ and $N$-point with $P$ parallelism, labeled as $\sigma_{N}^{s,k,P}$, ensures accurate data is provided to the total of $P$ MDCs. This process is performed in three transformations:
\begin{equation}
\sigma_{N}^{s,k,P}=\sigma_{N,3}^{s,k,P}\circ \sigma_{N,2}^{s,k,P}\circ\sigma_{N,1}^{s,k,P}
\end{equation}
The first permutation $\sigma_{N,1}^{s,k,P}$ is applied to serial bit dimension of ~(\ref{eq:block_p}) as:
\newcommand{\Cat}{\text{Cat}}
\newcommand{\Ra}[1]{R\{#1\}}
\begin{equation}{
\sigma_{N,1}^{s,k,P} =
\begin{cases}
\text{if } n - p - 1 \geq w_{N}^{s,k,P}: \\
\Cat([a_{n-p-1} : a_{w_{N}^{s,k,P}}], \Ra{[a_{w_{N}^{s,k,P}-1} : a_{0}]}) \\
\text{if } n - p - 1 < w_{N}^{s,k,P}: \\
\Ra{[a_{n-p-1} : a_{0}]} \\
\end{cases} \label{eq:first_p}
}\end{equation}

where the function $\Cat([u:v], [c:d])$ denotes the concatenation of two bit sequences: one ranging from $u$ to $v$, and the other from $c$ to $d$. The operation $R\{.\}$ denotes bit-reversal. The permutation $\sigma_{N,1}^{s,k,P}$ is defined to reverse the latter part of the serial dimension when the condition $n - p - 1 \geq w$ is fulfilled; otherwise, it reverses all the bits. This operation takes place in memory banks whose identifiers belong to $\{2P(s-1), \cdots, 2Ps\}$ at every stage.

\indent The mutable parameter $w_{N}^{s,k,P}$ indicates the number of bits that need to be reversed, with its value determined by $N$, $s$, and $k$. 
To explore the value of the parameter $w_{N}^{s,k,P}$ under various possibilities, we conduct mathematical modeling of the hardware behavior of the proposed FFT architecture through software programming. We summarize a generalized formula for calculating specific values, considering any computation stages of an arbitrary $N$-point executed on a $P$-parallelism radix-$2^k$ FFT processor condition. The formula is given as if $N > 2^k$:
\begin{equation}
{
w_{N}^{s,k,P}= 
\begin{cases}
\text{if } s \in \{1, S\}:\\
\indent \log_{2}N -1-\log_2 P  \\
\text{if } s \notin \{1, S\}:\\
 k \left( \left\lfloor \frac{\log_2 N}{k} \right\rfloor + s - S \right) +\\\log_2 N\%k-1-\log_2 P \\
\end{cases} \label{eq:casesreverse}
}
\end{equation}

\begin{figure}[ht!]
    \centering
    \includegraphics[width=0.4\textwidth]{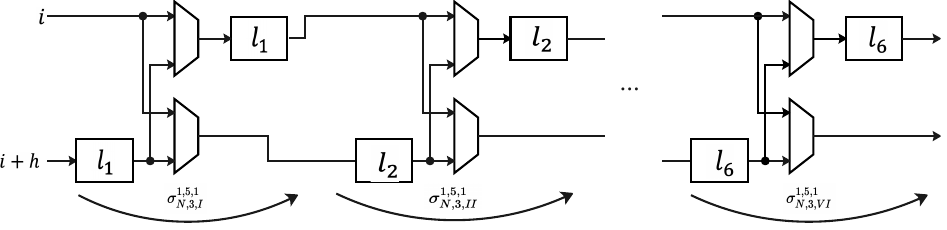}
    \caption{Circuit for a series of permutation}
    \label{fig:transformation}
\end{figure}

For example, 4096-point processed on the radix-$2^5$ processor of parallelism $P=1$ requires three stages, and the bit reverse number for each stage is obtained from ~(\ref{eq:casesreverse}) as $w_{4096}^{1,5,1}=11$, $w_{4096}^{2,5,1}=6$, $w_{4096}^{3,5,1}=11$. Therefore, in the first and last stages, the first permutation $(\sigma_{4096,1}^{s,5,1})$ requires reversing all serial bit dimensions in contrast to the second stage, where only the last six bits need to be reversed as calculated from~(\ref{eq:first_p}).\\
\indent The second permutation, $\sigma_{N,2}^{s,k,P}$, acts on the parallel dimension of (\ref{eq:block_p}). Consequently, it only alters the parallel branches after samples are read out, which refers to the Parallel Branch Change Circuit module in Fig.\ref{fig:overview}, redirecting the dataflow stream into different reshuffle circuit input terminals. The parallel permutation on $p=\log_2{2P}$ bits is given as:
\begin{equation}
\sigma_{N,2}^{s,k,P}=R\{[a_{n-1}:a_{n-p}]\}
\end{equation}
where denotes that reverse all bits in the parallel dimension. 
\indent The third permutation, $\sigma_{N,3}^{s,k,P}$, executed on both parallel and serial dimensions, is the final permutation before the MDC calculation. It facilitates fine coordination of indices within a block $B_{s}^{I}$. We introduce a parameter $h$ to determine that the third permutation takes place between the $i$-th and $(i+h)$-th parallel branches, where $h \in [0, P)$. Another parameter $l$ represents the length of delay FIFO. Permutation is achieved by interchanging the $l'=\log_2{l}$-th bit of the serial sequence with the $h'=\log_2{h}$-th bit of the parallel sequence as follows:
\begin{equation}
\begin{aligned}
\sigma_{N,3}^{s,k,P} &\left( a_{n-1}, \ldots, a_{n-p+h'}, a_{n-p} \mid a_{n-p-1}, \ldots, a_{l'}, a_{0} \right) \\
= &\left( a_{n-1}, \ldots, a_{l'}, a_{n-p} \mid a_{n-p-1}, \ldots, a_{n-p+h'}, a_{0} \right)
\end{aligned}
\end{equation}
which can also be represented as:
\begin{equation}
\sigma_{N,3}^{s,k,P} :  a_{n-p+h'} \leftrightarrow a_{l'}
\end{equation}
The hardware required to realize this permutation consists of two $l$-length FIFOs and two multiplexers, which refers to the Reshuffle Circuit module in Fig.\ref{fig:overview}. The two input terminals are connected to the $i$-th and $(i+h)$-th parallel branch as shown in Fig.\ref{fig:transformation}. However, a single $\sigma_{N,3}^{s,k,P}$ permutation does not suffice to transform $\sigma_{N,2}^{s,k,P}$ into $B_{s+1}^{I}$. Consequently, a series of permutations is used after $\sigma_{N,3}^{s,k,P}$.
We construct a mathematical model to determine the number of permutations in the composition and the precise values of $l$ and $h$ within each permutation, considering any computation stage of an arbitrary $N$-point executed on a $P$-parallelism radix-$2^k$ FFT processor. Our findings indicate that the number of third permutations can be taken from 0 to 6 times, denoted as $\sigma_{N,3,I}^{s,k,p} \ldots \sigma_{N,3,VI}^{s,k,p}  $ 
, which makes it extremely challenging to unify them into a general equation.

To name a few for better illustration, Tables \ref{your_table_label_I} and \ref{your_table_label_II} list a series of permutations at the first stage, with $1$-parallelism and $4$-parallelism on radix-$2^{5}$ for an arbitrary $N$-point, respectively. It is worth noting that as the parallelism of the FFT processor increases, the composition of the third permutation involves different permutation circuits, with the parameter $h$ varying among three distinct options (1, 2, 4). The third permutation is not a crucial operation for arbitrary points, as the symbol "$\backslash$" in both tables indicates that it should not be applied

\begin{table}[ht!]
\centering
\caption{Configuration of first stage with $1$-Parallelism radix-$2^{5}$}
\label{your_table_label_I}
\resizebox{0.45\textwidth}{!}{
\begin{tabular}{c|ll|ll|ll|ll|ll|ll}
\hline
\multirow{2}{*}{$n\%5$} & \multicolumn{2}{c|}{$\sigma_{N,3,I}^{1,5,1}$} & \multicolumn{2}{c|}{$\sigma_{N,3,II}^{1,5,1}$} & \multicolumn{2}{l|}{$\sigma_{N,3,III}^{1,5,1}$}  & \multicolumn{2}{l|}{$\sigma_{N,3,IV}^{1,5,1}$} & \multicolumn{2}{l|}{$\sigma_{N,3,V}^{1,5,1}$} & \multicolumn{2}{l}{$\sigma_{N,3,VI}^{1,5,1}$}\\ \cline{2-13}
 & \multicolumn{1}{c|}{$h$} &  \multicolumn{1}{c|}{$l$} & \multicolumn{1}{c|}{$h$} & \multicolumn{1}{c|}{$l$} &\multicolumn{1}{c|}{$h$} & \multicolumn{1}{c|}{$l$} & \multicolumn{1}{c|}{$h$} & \multicolumn{1}{c|}{$l$} & \multicolumn{1}{c|}{$h$} & \multicolumn{1}{l|}{$l$} &  \multicolumn{1}{c|}{$h$} & \multicolumn{1}{c}{$l$}\\ \hline\hline
0 & \multicolumn{1}{c|}{1} & \multicolumn{1}{c|}{8} & \multicolumn{1}{c|}{1} & \multicolumn{1}{c|}{1} & \multicolumn{1}{c|}{1}  & \multicolumn{1}{c|}{8} & \multicolumn{1}{c|}{1} & \multicolumn{1}{c|}{4} & \multicolumn{1}{c|}{1} & \multicolumn{1}{c|}{2}  & \multicolumn{1}{c|}{1} &\multicolumn{1}{c}{4} \\ \hline
1 & \multicolumn{1}{c|}{$\backslash$} & \multicolumn{1}{c|}{$\backslash$} & \multicolumn{1}{c|}{$\backslash$} & \multicolumn{1}{c|}{$\backslash$} & \multicolumn{1}{c|}{$\backslash$} & \multicolumn{1}{c|}{$\backslash$} & \multicolumn{1}{c|}{$\backslash$} & \multicolumn{1}{c|}{$\backslash$} & \multicolumn{1}{c|}{$\backslash$} & \multicolumn{1}{c|}{$\backslash$} & \multicolumn{1}{c|}{$\backslash$} & \multicolumn{1}{c}{$\backslash$} \\ \hline
2 & \multicolumn{1}{c|}{$\backslash$} & \multicolumn{1}{c|}{$\backslash$} & \multicolumn{1}{c|}{$\backslash$} & \multicolumn{1}{c|}{$\backslash$} & \multicolumn{1}{c|}{$\backslash$} & \multicolumn{1}{c|}{$\backslash$} & \multicolumn{1}{c|}{$\backslash$} & \multicolumn{1}{c|}{$\backslash$} & \multicolumn{1}{c|}{$\backslash$} & \multicolumn{1}{c|}{$\backslash$} & \multicolumn{1}{c|}{$\backslash$} & \multicolumn{1}{c}{$\backslash$} \\ \hline
3 & \multicolumn{1}{c|}{1} & \multicolumn{1}{c|}{2} & \multicolumn{1}{c|}{1} & \multicolumn{1}{c|}{1} & \multicolumn{1}{c|}{1} & \multicolumn{1}{c|}{2} & \multicolumn{1}{c|}{$\backslash$} & \multicolumn{1}{c|}{$\backslash$} & \multicolumn{1}{c|}{$\backslash$} & \multicolumn{1}{c|}{$\backslash$} & \multicolumn{1}{c|}{$\backslash$} & \multicolumn{1}{c}{$\backslash$} \\ \hline
4 & \multicolumn{1}{c|}{1} & \multicolumn{1}{c|}{4} & \multicolumn{1}{c|}{1} & \multicolumn{1}{c|}{1} & \multicolumn{1}{c|}{1} & \multicolumn{1}{c|}{4} & \multicolumn{1}{c|}{$\backslash$} & \multicolumn{1}{c|}{$\backslash$} & \multicolumn{1}{c|}{$\backslash$} & \multicolumn{1}{c|}{$\backslash$} & \multicolumn{1}{c|}{$\backslash$} & \multicolumn{1}{c}{$\backslash$} \\ \hline
\end{tabular}
}
\end{table}
\begin{table}[ht!]
\centering
\caption{Configuration of  first stage with $4$-Parallelism radix-$2^{5}$}
\label{your_table_label_II}
\resizebox{0.45\textwidth}{!}{
\begin{tabular}{c|ll|ll|ll|ll|ll|ll}
\hline
\multirow{2}{*}{$n\%5$} & \multicolumn{2}{c|}{$\sigma_{N,3,I}^{1,5,4}$} & \multicolumn{2}{c|}{$\sigma_{N,3,II}^{1,5,4}$} & \multicolumn{2}{l|}{$\sigma_{N,3,III}^{1,5,4}$}  & \multicolumn{2}{l|}{$\sigma_{N,3,IV}^{1,5,4}$} & \multicolumn{2}{l|}{$\sigma_{N,3,V}^{1,5,4}$} & \multicolumn{2}{l}{$\sigma_{N,3,VI}^{1,5,4}$}\\ \cline{2-13}
 & \multicolumn{1}{c|}{$h$} &  \multicolumn{1}{c|}{$l$} & \multicolumn{1}{c|}{$h$} & \multicolumn{1}{c|}{$l$} &\multicolumn{1}{c|}{$h$} & \multicolumn{1}{c|}{$l$} & \multicolumn{1}{c|}{$h$} & \multicolumn{1}{c|}{$l$} & \multicolumn{1}{c|}{$h$} & \multicolumn{1}{l|}{$l$} &  \multicolumn{1}{c|}{$h$} & \multicolumn{1}{c}{$l$}\\ \hline\hline
0 & \multicolumn{1}{c|}{4} & \multicolumn{1}{c|}{4} & \multicolumn{1}{c|}{2} & \multicolumn{1}{c|}{8} & \multicolumn{1}{c|}{1}  & \multicolumn{1}{c|}{2} & \multicolumn{1}{c|}{1} & \multicolumn{1}{c|}{1} & \multicolumn{1}{c|}{1} & \multicolumn{1}{c|}{2}  & \multicolumn{1}{c|}{$\backslash$} & \multicolumn{1}{c|}{$\backslash$} \\ \hline
1 & \multicolumn{1}{c|}{$\backslash$} & \multicolumn{1}{c|}{$\backslash$} & \multicolumn{1}{c|}{$\backslash$} & \multicolumn{1}{c|}{$\backslash$} & \multicolumn{1}{c|}{$\backslash$} & \multicolumn{1}{c|}{$\backslash$} & \multicolumn{1}{c|}{$\backslash$} & \multicolumn{1}{c|}{$\backslash$} & \multicolumn{1}{c|}{$\backslash$} & \multicolumn{1}{c|}{$\backslash$} & \multicolumn{1}{c|}{$\backslash$} & \multicolumn{1}{c}{$\backslash$} \\ \hline
2 & \multicolumn{1}{c|}{2} & \multicolumn{1}{c|}{1} & \multicolumn{1}{c|}{$\backslash$} & \multicolumn{1}{c|}{$\backslash$} & \multicolumn{1}{c|}{$\backslash$} & \multicolumn{1}{c|}{$\backslash$} & \multicolumn{1}{c|}{$\backslash$} & \multicolumn{1}{c|}{$\backslash$} & \multicolumn{1}{c|}{$\backslash$} & \multicolumn{1}{c|}{$\backslash$} & \multicolumn{1}{c|}{$\backslash$} & \multicolumn{1}{c}{$\backslash$} \\ \hline
3 & \multicolumn{1}{c|}{4} & \multicolumn{1}{c|}{1} & \multicolumn{1}{c|}{2} & \multicolumn{1}{c|}{2} & \multicolumn{1}{c|}{$\backslash$} & \multicolumn{1}{c|}{$\backslash$} & \multicolumn{1}{c|}{$\backslash$} & \multicolumn{1}{c|}{$\backslash$} & \multicolumn{1}{c|}{$\backslash$} & \multicolumn{1}{c|}{$\backslash$} & \multicolumn{1}{c|}{$\backslash$} & \multicolumn{1}{c}{$\backslash$} \\ \hline
4 & \multicolumn{1}{c|}{4} & \multicolumn{1}{c|}{2} & \multicolumn{1}{c|}{2} & \multicolumn{1}{c|}{4} & \multicolumn{1}{c|}{$\backslash$} & \multicolumn{1}{c|}{$\backslash$} & \multicolumn{1}{c|}{$\backslash$} & \multicolumn{1}{c|}{$\backslash$} & \multicolumn{1}{c|}{$\backslash$} & \multicolumn{1}{c|}{$\backslash$} & \multicolumn{1}{c|}{$\backslash$} & \multicolumn{1}{c}{$\backslash$} \\ \hline
\end{tabular}
}
\end{table}

In conclusion, when the proposed FFT architecture is configured to parallel mode, $\sigma_{N,1}^{s,k,P}$ is applied to read initial samples from memories according to the address generation specified Equation (\ref{eq:first_p}). Then, $\sigma_{N,2}^{s,k,P}$ is used to exchange the parallel branch. Following this, a composition of $\sigma_{N,3}^{s,k,P}$ is employed to adjust the indices order for MDC calculation. Finally, the output is delivered to neighboring memories in parallel architecture or primitive memories in memory-based architecture, repeating the aforementioned calculation pattern. In contrast, within a memory-based architecture, the primary distinction is that the dataflow back and forth within the same set of memory and MDC calculation resources.

\subsection{Conflict-Free Memory Access }\label{sec:sectionIII-C}
The proposed FFT design can be reconfigured into two computational modes: pipeline architecture and memory-based architecture. The conflict happens when old data is updated with new arrival data.In the first mode, intermediate results stored in memory are repeatedly overwritten with a different batch at the same stages. In contrast, in the memory-based architecture, intermediate results are repeatedly overwritten within the same batch across different iterations. Therefore, memory access patterns can be categorized into two scenarios.

Firstly, in the pipeline architecture case, the dataflow is as follows: the total number of $2P$ memories receive continuous real-time samples, which are then fed into a number of $P$ MDCs after the $\sigma_{N}^{1,k,p}$ permutation. Subsequently, the first intermediate result is streamed out to neighboring $2P$ memories. This dataflow pattern is consistently applied across all stages. Hence, memory access should be designed to prevent collisions.

From this discussion, the serial bit dimension in (\ref{eq:block_p}) is connected to the address in memory. Consequently, the bit permutation of the address is equal to $\sigma_{N,1}^{s,k,P}$, which is expressed as $\sigma_{mem} = \sigma_{N,1}^{s,k,P}$. The bit permutation of the read and write addresses can be derived from $\sigma_{mem}$ as:
\begin{equation}
\sigma_{mem}^{i} = \sigma_{R}^{i} \circ \sigma_{W}^{i}\label{eq:read_write}
\end{equation}
where $\sigma_{R}^{i}$ and $\sigma_{W}^{i}$ are the bit permutations of the circular counter used for generating the read and write addresses at the $i$-th batch. These permutations are defined as follows:
\begin{equation}
W_{A}^{i} = \sigma_{W}^{i}(c_{n-1}, c_{n-2}, \cdots, c_{1}, c_{0})\label{eq:write_batch}
\end{equation}
\begin{equation}
R_{A}^{i} = \sigma_{R}^{i}(c_{n-1}, c_{n-2}, \cdots, c_{1}, c_{0})\label{eq:read_batch}
\end{equation}
To guarantee conflict-free requirements, it must fulfill the principle that $(i+1)$-th batch of data is written in the same address as $i$-th batch of data has been read out\cite{garrido2020continuous}:
\begin{equation}
\sigma_{W}^{i} = \sigma_{R}^{i-1} \label{eq:read_write_batch}
\end{equation}
Given that the initial write address is the same as the circular counter, i.e., \( W_{A}^{i} = (c_{n-1}, c_{n-2}, \cdots, c_{1}, c_{0}) \), the permutation \(\sigma_{W}^{1}\) acts as an identity function. Consequently, \( R_{A}^{1} = \sigma_{N,1}^{s,k,P} \) is derived from (\ref{eq:read_write}). According to (\ref{eq:read_write_batch}) and (\ref{eq:write_batch}), \( W_{A}^{2} = R_{A}^{1} \). By applying (\ref{eq:read_write}), (\ref{eq:read_batch}), and (\ref{eq:read_write_batch}) again, it follows that \( R_{A}^{2} = Id = (c_{n-1}, c_{n-2}, \cdots, c_{1}, c_{0}) \) and \( W_{A}^{3} = R_{A}^{2} \). This establishes that two sets of addresses, specifically the natural order and the reversed order, access the memory in an interleaved manner.

Secondly, considering the memory-based architecture, data access only takes place in a total of $2P$ memories, and intermediate results directly return to the original $2P$ memories. Therefore, the first permutation $ \sigma_{N.1}^{s, k, P}$ referred to (\ref{eq:first_p}) isn't applied under this circumstance. When $N\in (2^{k},2^{3k}]$,nuance exists.In light of this fact, we introduce another bit permutation $\tilde{\sigma}_{N.1}^{s, k, P}$ ensuing after $\sigma_{N.1}^{s, k, P}$,which is defined as:
\begin{equation}
\tilde{\sigma}_{N.1}^{s, k, P}=\Cat\left(R\{[a_{n-p-1} : a_{n-p-\tilde{w}-1}]\}, [a_{n-p-\tilde{w}} : a_{0}]\right)
\end{equation}
where $\tilde{w} = \tilde{w}_{N}^{s, k, P}$. The permutation $\tilde{\sigma}_{N.1}^{s, k, P}$ represents the reversal of the initial segment of the serial dimension, which is the opposite operation of $\sigma_{N,1}^{s, k, P}$. Additionally, the original first permutation ($\sigma_{N.1}^{s, k, P}$) can be reformulated as a new composition of permutations($\hat\sigma_{N.1}^{s, k, P}$) which takes two steps:
\begin{equation}
\hat{\sigma}_{N.1}^{s, k, P} =\tilde{\sigma}_{N.1}^{s, k, P}\circ\sigma_{N.1}^{s, k, P}
\end{equation}
Note that the corresponding parameters of  $\tilde{w}_{N}^{s, k, P}$,${w}_{N}^{s, k, P}$ are varied in this context:
\begin{equation}
{w}_{N}^{s, k, P} = 
\begin{cases}
if  s\in\{1,2\}\land($N$\in (2^{2k},2^{3k}])\\
\indent \log_2 N - 1 - \log_2 P\\
if s = 3\land($N$\in (2^{2k},2^{3k}])\\
\indent k \left( \left\lfloor \frac{\log_2 N}{k} \right\rfloor -1  \right) + \log_2 N \% k - 1 - \log_2 P\\
if  (s = 1)\land($N$\in (2^{k},2^{2k}])\\
\indent \log_2 N - 1 - \log_2 P\\
if  (s = 2)\land($N$\in (2^{k},2^{2k}])\\
\indent 0
\end{cases}\label{eq:m_sigma}
\end{equation}
Comparing ${w}_{N}^{s, k, P}$ in (\ref{eq:m_sigma}) with (\ref{eq:casesreverse}), it is noted that only the first stage of the first permutation is identical, and another parameter $\tilde{w}_{N.1}^{s, k, P}$ is given as:
\begin{equation}
\tilde{w}_{N}^{s, k, P} = 
\begin{cases}
if  (s = 2)\land($N$\in (2^{2k},2^{3k}])\\
\indent k \left( \left\lfloor \frac{\log_2 N}{k} \right\rfloor -1  \right) + \log_2 N \% k - 1 - \log_2 P\\
else \\
\indent 0\label{eq:m_sigma_1}
\end{cases}
\end{equation}
\indent Observing equations (\ref{eq:m_sigma}) and (\ref{eq:m_sigma_1}), we notice that the permutation $\tilde{\sigma}_{N.1}^{s, k, P}$ is specifically adapted to $N$-point, which requires three iterations. For example, consider using a radix-$2^{5}$ memory-based architecture to process 4096 points with $P = 4$ parallelism. The bit reverse number for each iteration is ${w}_{4096}^{1, 5, 4} = 9$, ${w}_{4096}^{2, 5, 4} = 9$, $\tilde{w}_{4096}^{2, 5, 4} = 4$, and ${w}_{4096}^{3, 5, 4} = 4$, respectively. Consequently, during the first iteration, all serial bits are reversed. All bits are reversed in the second iteration, followed by reversing the initial 4 bits. In the final iteration, only the last 4 bits require reversal.

Due to \(\hat{\sigma}_{N.1}^{s, k, P} = \sigma_{mem}\), the generation of read/write addresses can refer to the previous analysis, where
\begin{equation}
\sigma_{mem}^{s} = \sigma_{R}^{s} \circ \sigma_{W}^{s}\label{eq:sd}
\end{equation}

During the initial stage, samples are stored in natural counter-order; therefore, according to equations (\ref{eq:sd}) and (\ref{eq:read_write_batch}), \(\sigma_{R}^{1} = \hat{\sigma}_{N,1}^{1,k,P}\) and \(\sigma_{W}^{2} = \sigma_{R}^{1}\) are obtained. Consequently, using equation (\ref{eq:sd}),
\begin{equation}
\sigma_{R}^{2} = \hat{\sigma}_{N,1}^{2,k,P} \circ \sigma_{W}^{-1 2}
\end{equation}
is derived, where \(\sigma_{W}^{-1 2}\) is the inverse operation of bits all reversal, i.e., \(\sigma_{W}^{-1 2} = Id\). Thus, it is concluded that \(\sigma_{R}^{2} = \hat{\sigma}_{N,1}^{2,k,P}\) , which denotes the bits permutation pattern of read counter in stage 2 is equal to \(\hat{\sigma}_{N,1}^{2,k,P}\).The subsequent deduction is omitted, but the analysis is equivalent to the previous one.\\
\indent In conclusion, we introduce two types of conflict-free memory access: parallel mode and memory-based mode. The former is suitable for any $N$-point with the straightforward use of two sets of interleaved address patterns. Nevertheless, the latter is restricted to $N\in(2^{2k},2^{3k}]$  due to the requirement for in-place memory replacement.

\section{Comparison and Experiment Results}\label{sec:sectionV}
\subsection{Theoretical Comparison}
In this section, we will separately compare the proposed architecture operating in different modes with the previous architecture. 
First and foremost, Table \ref{your_table_label_III} lists other relevant memory-based architectures, all of which require memory with a range of $N$ addresses for controlling the variable. Compared to \cite{tsai2011generalized}, \cite{kaya2023memory}, and \cite{wang2020scheduling}, the proposed architecture can handle FFT processing at a higher radix, specifically up to radix-$2^5$, which benefits the processing of large size sets. In contrast, other mixed-radix MDC architectures, such as those in \cite{tsai2011generalized} and \cite{wang2020scheduling}, are limited to radix-$2^3$. Likewise, our proposed work aims to support FFT lengths ranging from 32 to 512K points, unlike the architectures presented in \cite{tsai2011generalized} and \cite{wang2020scheduling}. This significantly improves the adaptivity for handling large amounts of data, making it suitable for data-intensive applications.

Consequently, the number of iterations is drastically reduced compared to the previous approach by 40$\%$ or more, as it is calculated with respect to the radix and FFT length. In addition, our proposed work demonstrates a clear advantage in terms of parallelism, leading to higher overall throughput as the samples can be processed simultaneously with 8 branches.

Overall, given that the cycles per iteration are approximately equivalent to  $N/P_{c}$, the processing time can be calculated as $ \textit{iteration number} \times \textit{cycles per iteration}$. When processing the same points, our proposed work outperforms the previously existing architecture, resulting in a 70$\%$ or more reduction in execution time due to the higher radix and parallelism configuration.

\begin{table}[ht!]
\caption{Comparison of Memory-Based FFT Processor }
\label{your_table_label_III}
\resizebox{0.45\textwidth}{!}{
\begin{tabular}{c|c|c|c|c}
\hline
                & Tsai'11\cite{tsai2011generalized}     & Kaya'23\cite{kaya2023memory}     & Wang'20\cite{wang2020scheduling}       & This Work         \\ \hline\hline
Radix           & radix-$2/2^{2}/2^{3}$ & radix-2     & radix-$2/2^{2}/2^{3}$ & radix-$2/2^{2}/2^{3}/2^{4}/2^{5}$ \\ \hline
FFT Length     & 64$\sim$4096     &  N        & 2048$\sim$16K    & 32$\sim$32K          \\\hline
Parallelism     & $P_{c}=2$        & $P_{c}=2$  & $P_{c}=4$  & $P_{c}=4$           \\ \hline
Mem banks       & 4           & 4           & 8           & 8                 \\ \hline
Mem size       & N           & N          & N           & N                \\ \hline
Iterations      & $\lceil {n}/{3}\rceil$       & $\lceil {n}/{2}\rceil$       & $\lceil {n}/{3}\rceil$      & $\lceil {n}/{5}\rceil$             \\ \hline
Processing Time & $\lceil {n}/{3}\rceil N/P_{c}$       &$\lceil {n}/{2}\rceil N/P_{c}$         &$\lceil {n}/{3}\rceil N/P_{c}$             & $\lceil {n}/{5}\rceil N/P_{c}$\\ \hline
\end{tabular}
}
\end{table}

Secondly, comparing our proposed architecture with a pipelined approach, Table \ref{your_table_label_IV} delineates their differences. The average utilization, denoting the mean ratio of active butterfly units across various points, serves as a metric for evaluation. To ensure fairness in assessment, we assume an equal allocation of hardware resources for computing 512k points. 

Twenty butterfly units are employed in the proposed radix-$2^5$ architecture, with one unit remaining idle. This configuration is aligned with the architecture proposed by \cite{Garrido2013}, which operates under resource constraints and supports only a single level of parallelism. Consequently, a significant portion of the hardware remains idle during small-size FFT operations.In contrast, our architecture enhances utilization by leveraging increased parallelism. We achieve double or quadruple utilization rates by employing four sets of identical MDCs simultaneously. This adaptive hybrid FFT architecture significantly optimizes hardware utilization.

\subsection{Post-Implementation Results}
The architecture presented has been implemented by Chisel, which has been programmed to Virtex UltraScale+ VCU118(xcvu9p-flga2104-2L-e)FPGA. The design compiler is Xilinx Vivado 2019.2.The memories for transferring and storing data are implemented using ultra RAMs(URAMs). Each memory has 256K addresses of 32 bits for the real or imaginary part, leading to a total of 128Mb URAM usage. Meanwhile, block RAMs(BRAMs) are used enough to store twiddle factors including sine and cosine coefficients.
\begin{table}[ht!]
\centering
\caption{Comparison of Pipelined FFT Processor}
\label{your_table_label_IV}
\renewcommand{\arraystretch}{1.0} % 调整表格的高度
\resizebox{0.45\textwidth}{!}{
\begin{tabular}{c|c|c|c|c}
\hline
                   & Radix              & Parallelsim & FFT length & Avg Utilization \\ \hline\hline
\multirow{3}{*}{This work} & \multirow{3}{*}{radix-$2^5$} & 1           & 2048$\sim$512K     & 75$\%$              \\ \cline{3-5} 
                   &                    & 2           & 64$\sim$1024      &  80$\%$               \\ \cline{3-5} 
                   &                    & 4           & 2$\sim$32    &    60$\%$              \\ \hline
\multirow{3}{*}{Garrido'13 \cite{Garrido2013}} & \multirow{3}{*}{radix-$2^5$} & 1           & 2048$\sim$512K     & 75$\%$               \\ \cline{3-5} 
                   &                    & 1           & 64$\sim$1024      & 40$\%$            \\ \cline{3-5} 
                   &                    & 1           & 2$\sim$32     & 15$\%$              \\ \hline
\end{tabular}
}
\end{table}

\begin{table}[!t]
\centering
\caption{Implementation Results of Proposed FFT Processors}
\label{your_table_label_V}
\resizebox{0.5\textwidth}{!}{
\renewcommand{\arraystretch}{1.2} % 调整表格行高
\begin{scriptsize} % 调整字体大小
\begin{tabular}{|c|c|c|c|c|c|c|}
\hline
 & Num. of DSP48E2s & Num. of LUTs & Num. of FFs \\ \hline
 Proposed Work & 45365       & 76183        &1500  \\ \hline
 & Block RAMs (36K-bit) & Ultra RAMs & Freq. (MHz) \\ \hline
 Proposed Work & 444                              & 768        & 196.8   \\ \hline
\end{tabular}
\end{scriptsize}
}
\end{table}

\indent The detailed results of the FPGA implementation are provided in Table \ref{your_table_label_V} using $N = 512K,P=1$ as an example. The constant complex multipliers and non-trivial complex multipliers are implemented using DSP48E2 cells, where the complex multiplier consumes sixteen DSP48E2 cells. The external permutation circuits for data access are realized using LUT and FF slices. The proposed work requires more LUT and FF resources for the commutators to ensure accurate computing results. According to Table \ref{your_table_label_V}, the proposed architecture can achieve a frequency of up to 196.8 MHz.

\section{Conclusion}\label{sec:sectionVI}
This paper introduces an adaptive hybrid FFT processor with radix-$2^5$ multi-path delay commutators (MDC). Supporting FFT lengths up to 512K points, it improves hardware utilization, and reduces computational cycles, offering a flexible, high-performance solution for large-scale FFT applications.
% Bibliography

\end{document}